# Phonon instability and structural phase transitions in Vanadium under High pressure

Ashok K. Verma and P. Modak

*High Pressure Physics Division, Bhabha Atomic Research Centre,*
*Trombay, Mumbai-400085, India*

## Abstract

Results of the first-principles calculations are presented for the group-VB metals V, Nb and Ta up to couple of megabar pressure. An unique structural phase transition sequence $BCC \xrightarrow{\sim 60 GPa} R\text{hom}1(\alpha = 110.50^{\circ}) \xrightarrow{\sim 160 GPa} R\text{hom}2(\alpha = 108.50^{\circ}) \xrightarrow{\sim 434 GPa} BCC$ is predicted in V. We also find that BCC-V becomes mechanically and vibrationally unstable at around 112 GPa pressure. Similar transitions are absent in Nb and Ta.

**Introduction:**

Group-VB transition metals V, Nb, and Ta crystallizes in body-centered cubic (BCC) structure at ambient pressure and temperature conditions. These metals are of great use due to their high thermal, mechanical and chemical stabilities. Recently these metals have been the subject of numerous experimental and theoretical studies [1-7] in Mbar pressure regions. The Nb is known to have the highest superconducting transition temperature ($T_c$) among elemental solids at ambient pressure [1]. The Ta is used in high-pressure experiments as a pressure standard. Experimentally it is known that V, Nb and Ta remain stable in the BCC structure at least up to 150 GPa [2,4] pressure.

However, the linear response phonon calculations using full potential linear muffin-in orbital (FP-LMTO) method of V by Suzuki and Otani had shown the softening of the transverse acoustics (TA) phonon mode at ~120 GPa pressure, which eventually becomes imaginary at pressures higher than 130 GPa [5]. The subsequent work of Landa et al. using exact muffin-tin orbital (EMTO) method had also shown the anomalous pressure behavior of the $C_{44}$ elastic constant. They found that the $C_{44}$ starts softening above 50 GPa pressure and drops to zero value at about 180 GPa pressure. In the pressure range of 180-270 GPa the value of $C_{44}$ is negative. Above to this, the $C_{44}$ adopts the usual pressure behavior. These are an indication of the development of a structural instability in BCC lattice. However, these authors corroborated this behavior to the Fermi-surface nesting and the possibility of any structural phase transition was not explored [6].

In this paper we are reporting our ab-initio density functional theory based results for the V, Nb and Ta under high pressure. For V, BCC to rhombohedral structural phase transition has been predicted at around 60 GPa pressure and it remains in rhombohedral structure at least up to 332 GPa pressure. However, at 434 GPa pressure the BCC structure re-appears in our calculations. This behavior is not found in the Nb and Ta.. But lattice expansion in Nb resulted a new minimum around 111° which is lower in energy relative to BCC. By analogy, we expect such features in Ta will appear even at higher lattice expansions.

**Computational method:**

The cubic structures namely simple cubic, body centered cubic (BCC) and face-centered cubic (FCC) crystal structures are related to a rhombohedral structure. The simple cube can be viewed as a rhombohedron with angle ($\alpha_{rhom}$) equal to 90°. The primitive cells of BCC and FCC are rhombohedron with angle $\alpha_{rhom}$ equal to 109.47° and 60° respectively [8]. Thus it is possible to generate all theses cubic structures from a rhombohedron with atom at (0,0,0) by varying the angle. Here, we have performed total energy calculations taking a rhombohedral unit cell for V, Nb and Ta as a function of $\alpha_{rhom}$ in the range of 54°-112° at several volumes. All the calculations for V were done using the pseudopotential based PWSCF computer code [9]. The total energy calculations are based on density functional theory and the phonon calculations on density functional perturbation theory [10]. An ultrasoft pseudopotential for V was used with 40 Ry plane wave energy cut off and a 400 Ry cut off in the expansion of the augmentation charges. Phonon dynamical matrices were computed on a uniform 6**x**6**x**6 grid of **q**-points in the BZ of BCC structure. This leads to total 16 **q**-points in the irreducible BZ. Using Fourier interpolation, we then constructed and diagonalized the dynamical matrices on a denser grid (24**x**24**x**24). The calculations for the Ta and Nb were performed using the VASP computer program with PAW pseudopotentials taking  an energy cut off of 400 eV for plane wave expansions [11-13]. Results were cross checked with both programs at a few volumes for V and Nb. The *s* and *p* semicore levels of respective elements were included in the valence states. For the exchange-correlation term the generalized gradient correction approximation of Perdew-Buke-Ernzerof (GGA-PBE) [14] was used. A 18**x**18**x**18 **k**-point mesh for the Brillouin zone (BZ) integration was used for all the total energy calculations. The energy convergence with respect to computational parameters was carefully examined.

**Results and discussion:**

The zero-pressure and zero-temperature results for lattice constant $a_0$, bulk modulus $B_0$, and the first pressure derivative of the bulk modulus *B'* for V, Ta, and Nb are obtained by fitting the total energies versus lattice constant data to a fourth order polynomial. Calculated equilibrium lattice constants are in good agreement with the experimental values (within the error of 1%). Bulk modulus and its first pressure derivates are also in good agreement with existing experimental data (see table for comparison). Calculated pressure-volume relations matches well with existing experimental data and shown in Fig.1. Hence justifies the use of respective presudopotentials and other computational parameters.

Table-1: The experimental data shown in brackets by symbol (*) is of Takemura [2] and by symbol (+) is of Dewaele et. al.[7].

| Element | Lattice constant ($a_0$) | Bulk Modulus ($B_0$) | $B'$ |
|---|---|---|---|
| V | 2.998 (3.00)* | 182 (165,188)* | 3.70 (3.5,4.04)* |
| Nb | 3.309 (3.30)* | 172 (153,168)* | 3.40 (3.9,2.2)* |
| Ta | 3.320 (3.30)+ | 208 (194,)+ | 3.13 (3.25)+ |

Fig.2 depicts our calculated total energy as a function of angle $\alpha_{rhom}$ for V at equilibrium volume. The SC is on the absolute maximum and FCC is on the local maximum indicating the unstable nature of theses structures for a given rhombohedral distortion. The BCC is in the absolute minimum of the total energy curve as per expectation. However, there are two local minima around angles 56° and 68° (see the inset of Fig.2). To our best of knowledge no body has shown that for V, FCC and SC structures are mechanically unstable. The energy variation in the range of angles 106° -111° is about 4.0 mRy per atom (see Fig.3a). Similar calculations were performed at many volumes covering more than 400 GPa pressures. The Figs.3(a, b) show total energy variation as a function of angle $\alpha_{rhom}$ in the range of 106°-112° at several pressures. The reason for showing the energy variation only in this range of angles is two fold. Firstly, BCC is of lowest energy among all structures and secondly no extra feature appear around the FCC except that local minima become more pronounced with few degree shift. Particularly at 112 GPa pressure the minimum around 68° is shifted to angle 70° and the minimum around 56° shifted to 57° . At this pressure 70° minimum is lower in energy by 7 mRy per atom relative to the 57° minimum and around 12 mRy per atom relative to FCC. However, SC remains on the absolute maximum of the total energy curve up to 400 GPa pressure, ruling out the possibility of BCC to SC transition as predicted earlier [15].

Below 60 GPa pressure the behavior of energy curves remain unchanged. However at 60 GPa a new minimum appear at angle 110.50° (say, Min1) leading to a structural phase transition. Now BCC energy is higher by amount of 0.05 mRy per atom relative to Min1. Further increase in pressure leads to development of another minimum at angle108.50° (say, Min2) and BCC lies on a local maximum. At 112 GPa pressure energy difference between Min1 and Min2 is about –0.057 mRy per atom and that between Min1 and BCC is -0.10 mRy per atom. Thus now BCC-V becomes mechanically unstable toward a rhombohedral distortion which can lead to softening of $C_{44}$ elastic constant as it related to trigonal shear. In fact Land et al [6] predicted $C_{44}$ softening in the same pressure region. It is to be noted that in our work the BCC to rhombohedral structural phase transition has occurred at much lower pressure (~ 60 GPa). These energy changes are found to persist even with denser (*e.g.* 28**x**28**x**28 **k**-point) mesh for the BZ integration and with larger plane wave energy cut offs (e.g., 65 Ry). Thus it confirms that this feature is not an artifact of calculation but it is real. In fact, as these calculations were performed with same rhombohedral cell for all values of angles; inaccuracies due to computational parameters will be same. At 112 GPa pressure the calculations were also repeated with VASP using PAW pseudopotential [11-13] but results were invariant.

In our calculations the frequency of TA mode becomes imaginary at some **q**-values along the **Γ-H** direction of the BZ (Fig. 3c) at 112 GPa pressure; lower compared to ~130 GPa pressure reported by Suzuki and Otani [5]. The reason of this mismatch of pressure can be due to the difference in computational techniques and approximations used for exchange-correlation terms.

The total energy behavior between angles 106°-112° at 160 GPa is similar to that at 112 GPa except now the Min2 becomes lower in energy relative to Min1 by 0.20 mRy per atom leading to an iso-structural phase transition. Now Min2 energy is lower by 0.14 mRy per atom relative to BCC. Further increase in pressure lead to disappearance of Min1 and the energy difference between Min2 and Min1 at 240 GPa is around -0.55 mRy per atom. However, the energy difference between the BCC and Min1 is 0.14 mRy per atom. At 332 GPa, the Min2 is lower in energy by 0.019 mRy per atom relative to 109.47° (BCC) energy value. Further increase in pressure lead to disappearance of the Min2 also and at 434 GPa pressure there is only one minimum located at 109.47°. In our phonon calculations at 160 GPa pressure the TA mode frequency becomes positive and further increase in pressure lead to usual pressure behavior, *i.e*, hardening with pressure [Fig. 3(c)]. The calculations for pressures higher than 300 GPa were checked with VASP code using PAW pseudopotential including the nonlinear core corrections which are supposed to be taken care the core-core overlap at higher pressures. Our results remain unchanged.

We have also performed total energy calculations for hexagonal closed packed (HCP) lattice at optimized *c/a* axial ratio (1.829) at two pressures; one at ambient and other at 332 GPa pressure and its energy is found always higher compared to BCC. The energy difference (19.6 mRy/atom at ambient pressure) between BCC and HCP increases with pressure (40 mRy/atom at 332 GPa ) thus rules out the BCC-HCP transition.

It is to be noted that work reported by Ding et. al. had claim the observation of BCC to rhombohedral transition near 69 GPa pressure [17] which also supports our predictions.

Similar calculations were repeated for the Nb and Ta and the results of total energy variations with $\alpha_{rhom}$ at several pressures are shown in Figs. 4(a,b). No new features appear in total energy curves under pressure like those in V and thus no phonon calculations were attempted. These are in agreement with earlier elastic constant and phonon calculations [3,6,16]. The only known anomaly for these two elements in the $C_{44}$ is its slope change with pressure, which occurs at 40 GPa for Nb [6] and in 100 GPa region for Ta [16]. But lattice expansion in Nb resulted a new minimum around 111° which is lower in energy relative to BCC by 11 meV per atom at 6.7% lattice expansion. By analogy, we expect such features in Ta will appear even at larger lattice expansions.

**Conclusion:**

In conclusion, high pressure structural behavior of BCC metals of group-VB are investigated using density functional theory. Calculations for V show the onset of a phonon softening and a rhombohedral instability at 60 GPa pressure which may lead to $C_{44}$ elastic constant softening as predicted earlier [6] in the same pressure region. Thus, we predict a BCC to rhombohedral structural phase transition ($\alpha_{rhom}$=110.5°) at around 60 GPa pressure in V. It exists in rhombohedral structure at least up to 330 GPa pressure, however its angle is changed to 108.50° at 160 GPa pressure resulting to an iso-structural phase transition. Finally at around 434 GPa pressure, it transforms to BCC structure again. This behavior is not found in the Nb and Ta.

**References:**

[1] V. V. Struzhkin, Y. A. Timofeev, R. J. Hemley and H. K. Mao, Phys. Rev. Lett. **79**, 4262 (1197).

[2] K. Takemura, '*Porc. Int. Conf. On High Pressure Science and Technology, AIRAPT-17*' (Honolulu, 1999), (Sci. Technol. High Pressure vol 1) *ed*. M. H. Maghnani, W. J. Nellis and M. F. Nicol (India: University Press) page 443 (2000).

[3] J. S. Tse, Z. Li, K. Uehara, Y. Ma and R. Ahuja, Phys. Rev. B **69**, 132101 (2004).

[4] H. Cynn and C. S. Yoo, Phys. Rev. B **59**, 8526 (1999).

[5] N. Suzuki and M. Otani, J. Phys.:Condens. Matter **14**, 10869 (2002).

[6] A. Landa, J. Klepeis, P. Soderlind, I. Naumov, O. Velikokhatnyi, L. Vitos and A. Ruban, J. Phys.:Condens. Matter **18**, 5079 (2006).

[7] A. Dewaele, P. Loubeyre and M. Mezouar, Phys. Rev. B **70**, 094112 (2004).

[8] C. Kittel, '*Introduction to Solid State Physics*' Fifth Edition (1993).

[9] Website, *http://www.pwscf.org*

[10] S. Baroni, S. de Gironcoli, A. D. Corso and P. Giannozzi, Rev. Mod. Phys. **73**, 515 (2001).

[11] G. Kresse and J. Furthmuller, Phys. Rev. B. **54**, 11169 (1996).

[12] G. Kresse and J. Hafner, Phys. Rev. B **47**, 558 (1993).

[13] G. Kresse and J. Joubert, Phys. Rev. B **59**, 1758 (1999).

[14] J. P. Perdew, K Burke and, M Ernzerhof, Phys. Rev. Lett. **77,** 3865 (1996).

[15] C. Nirmala Louis and K. Iyakutti, Phys. Rev. B **67**, 094509 (2003).

[16] O. Gulseren and R. E. Cohen, Phys. Rev. B **65**, 064103 (2002).

[17] Ding et. al., Phys. Rev. Lett. **98,** 085502 (2007).

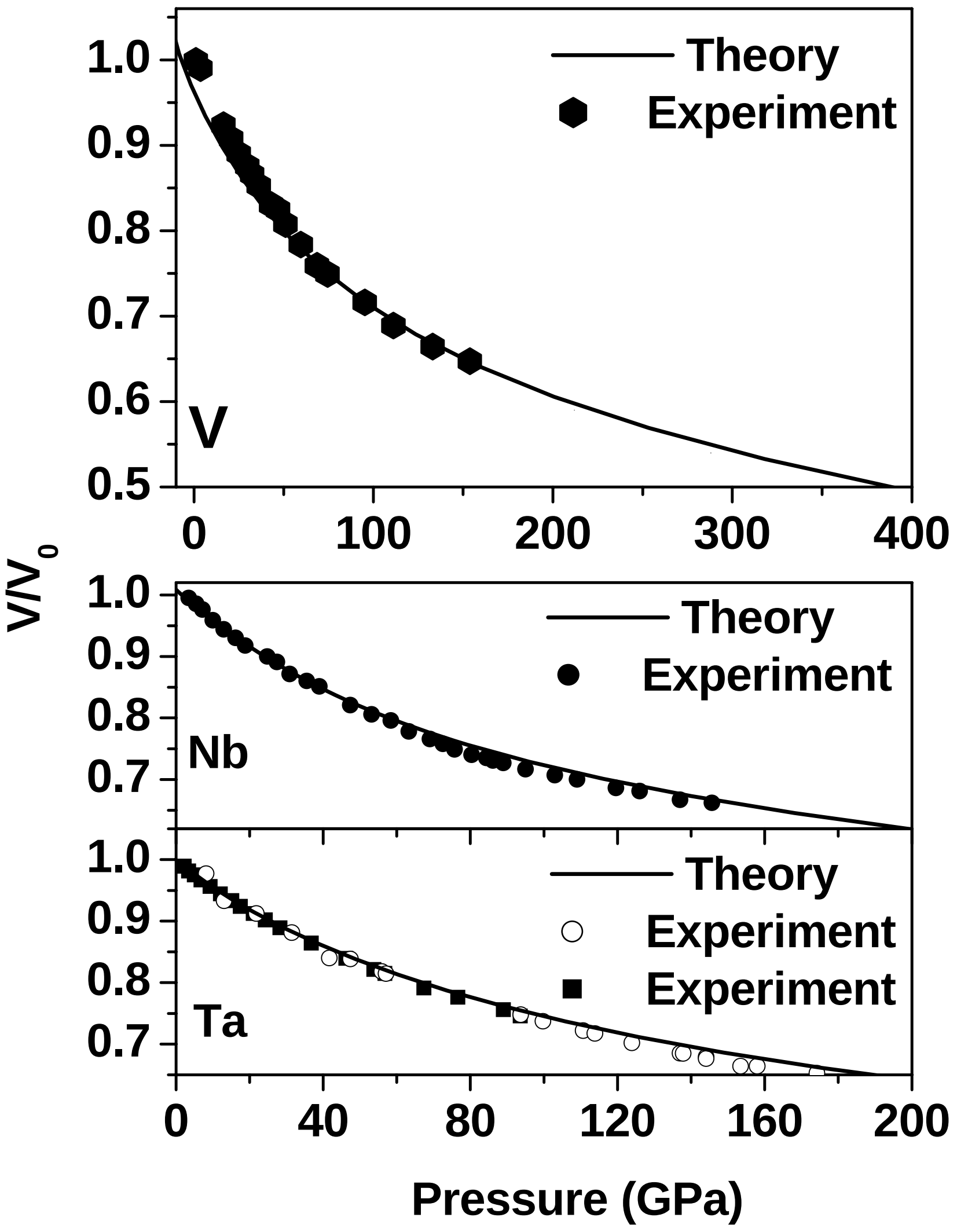

**Fig. 1:** The calculated P-V relations are shown with solid lines. The experimental data for V and Nb are from Ref. [2]. The experimental data for Ta shown by symbol (O) is from Ref. [4] and other is form Ref. [7]. The $V_0$'s are respective theoretical and experimental equilibrium volumes.

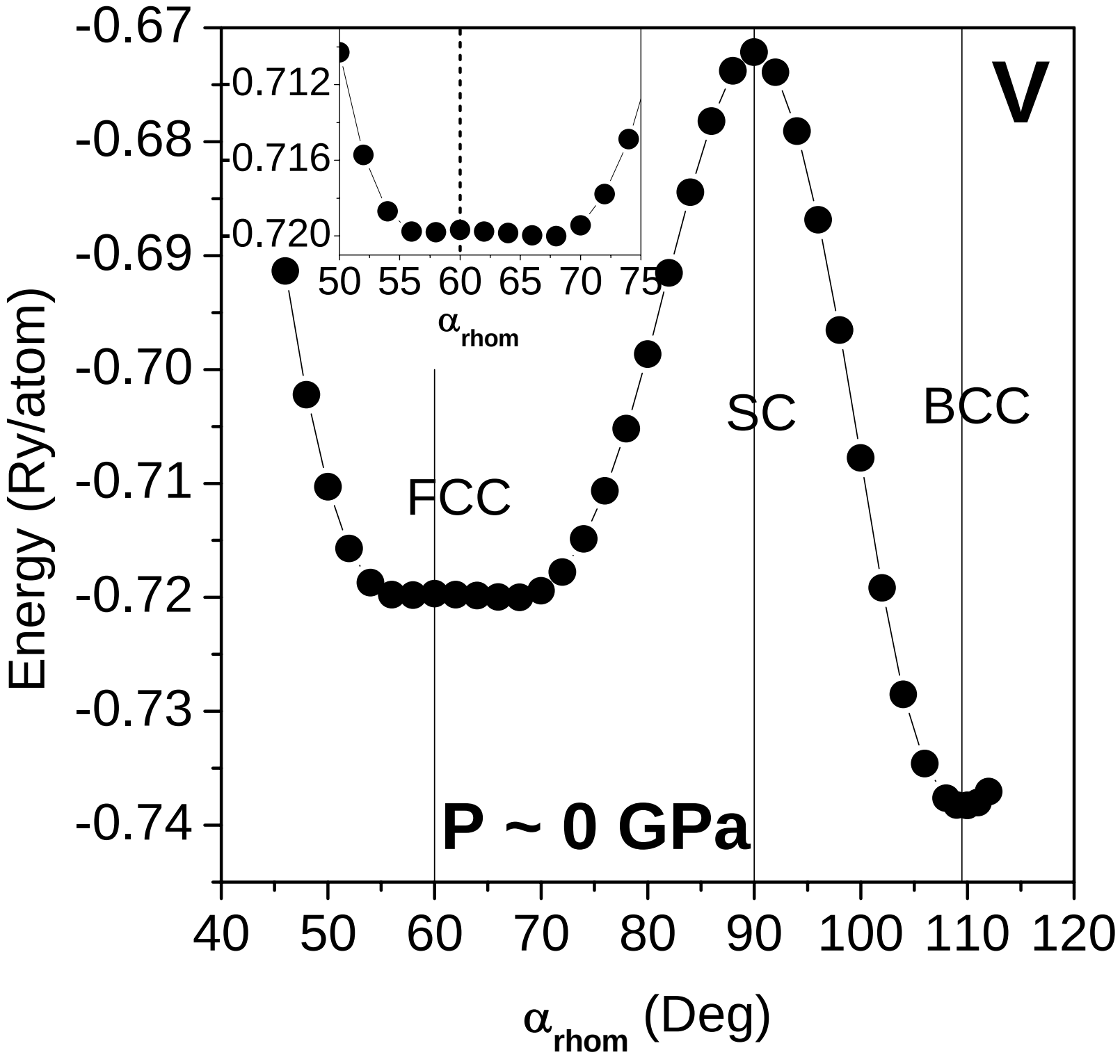

**Fig. 2:** Total energy variation with $\alpha_{rhom}$ at volume 13.48 Å$^3$ per atom.

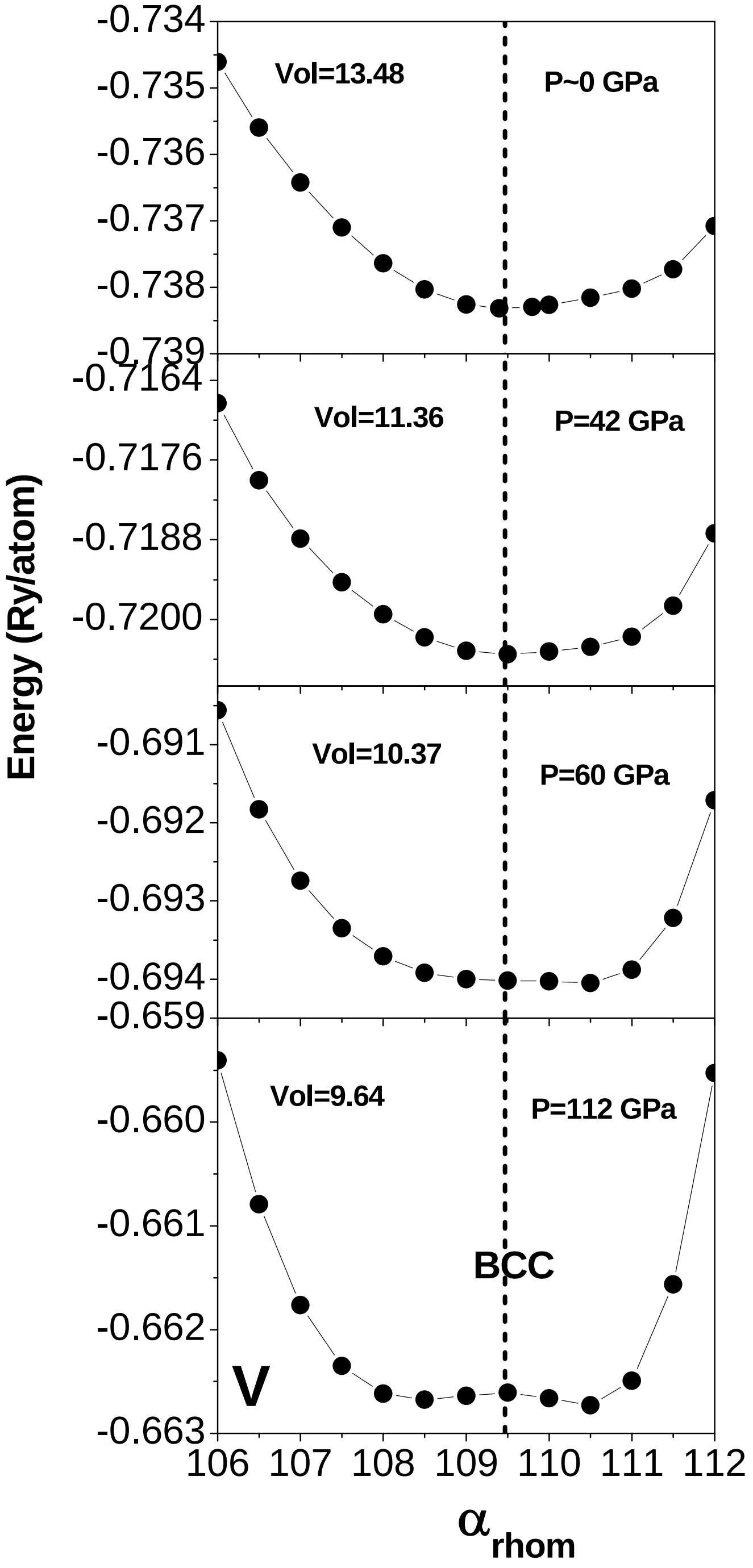

**Fig. 3(a)**

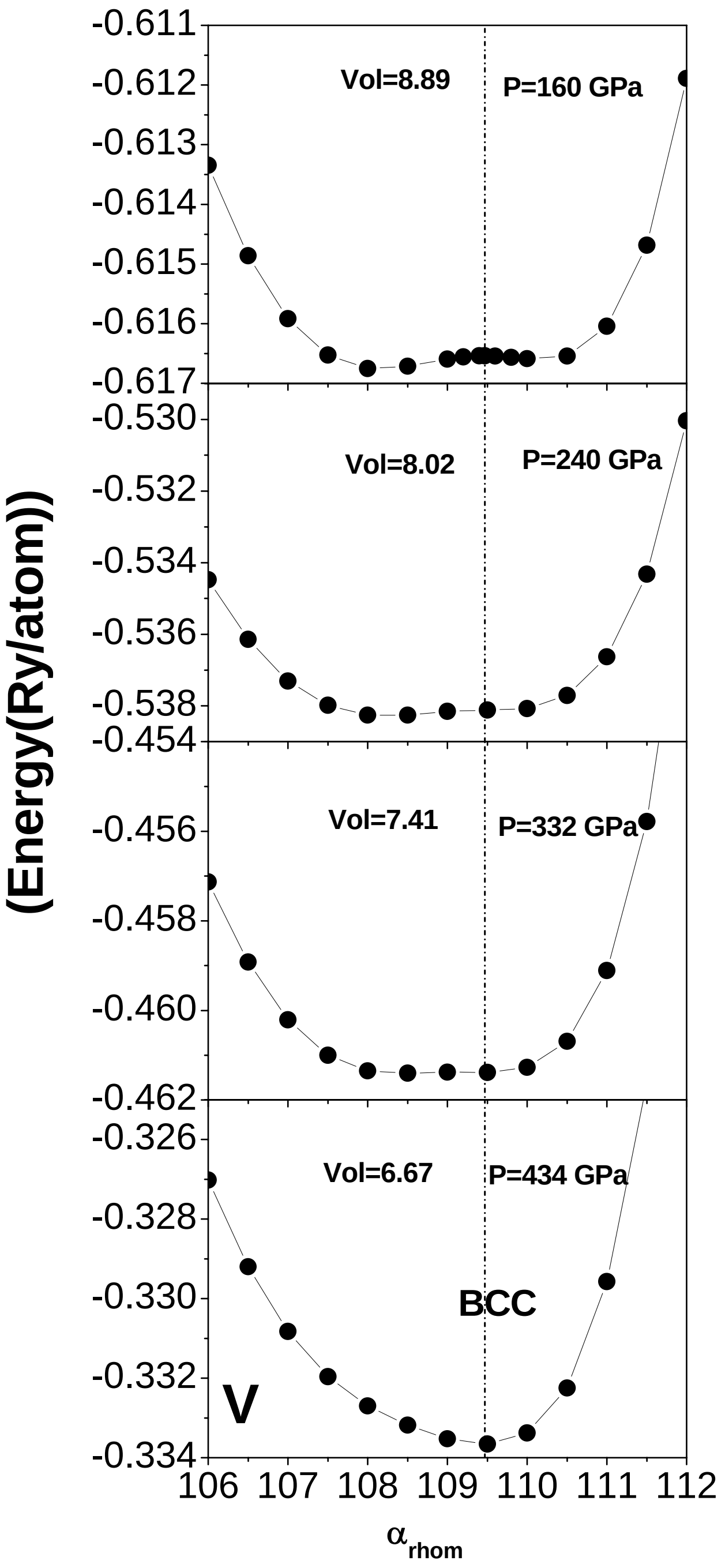

**Fig. 3(b)**

**Fig. 3(a,b):** The total energy variation with angle $\alpha_{rhom}$ at different pressures for V. All the volumes are in unit of Å$^3$ per atom.

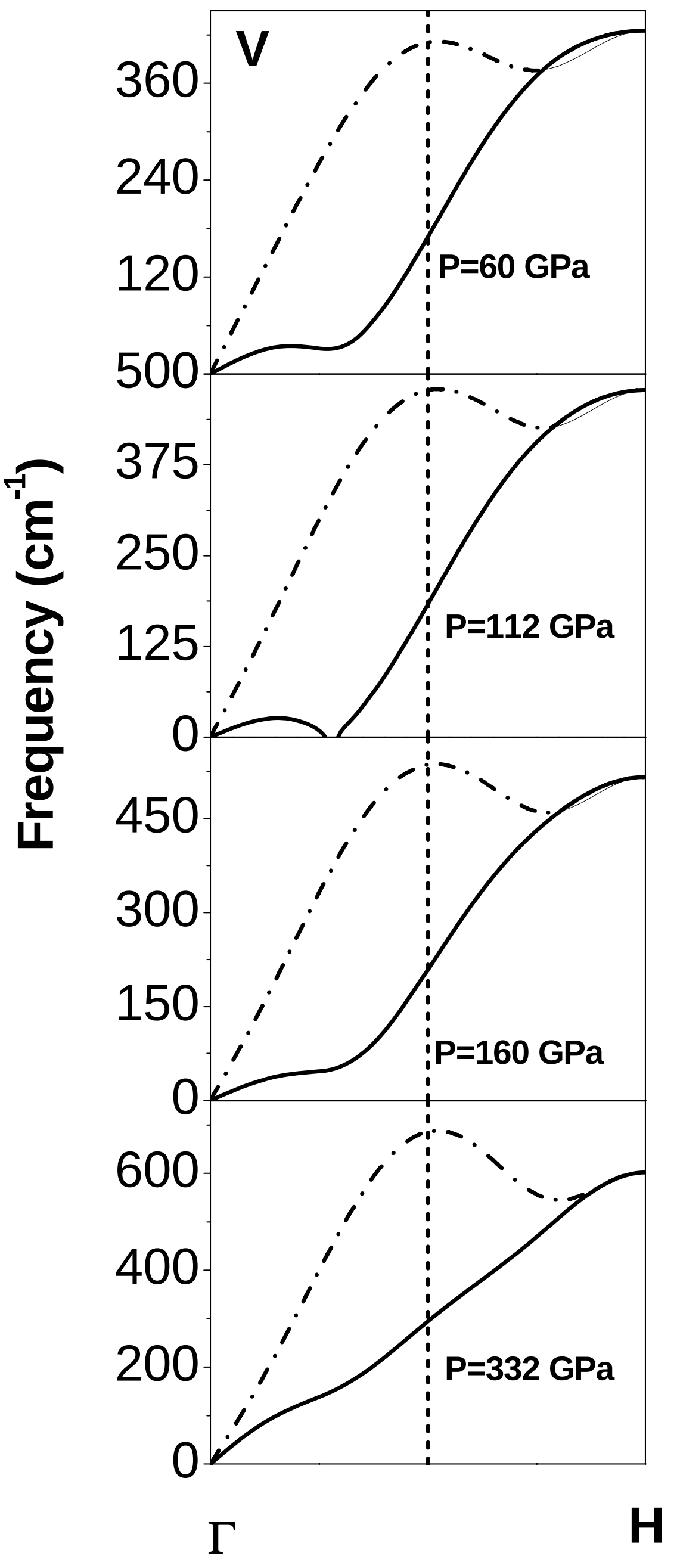

**Fig. 3c**: The pressure variation of the phonon modes frequency of V along **Γ-H** direction of the BZ.

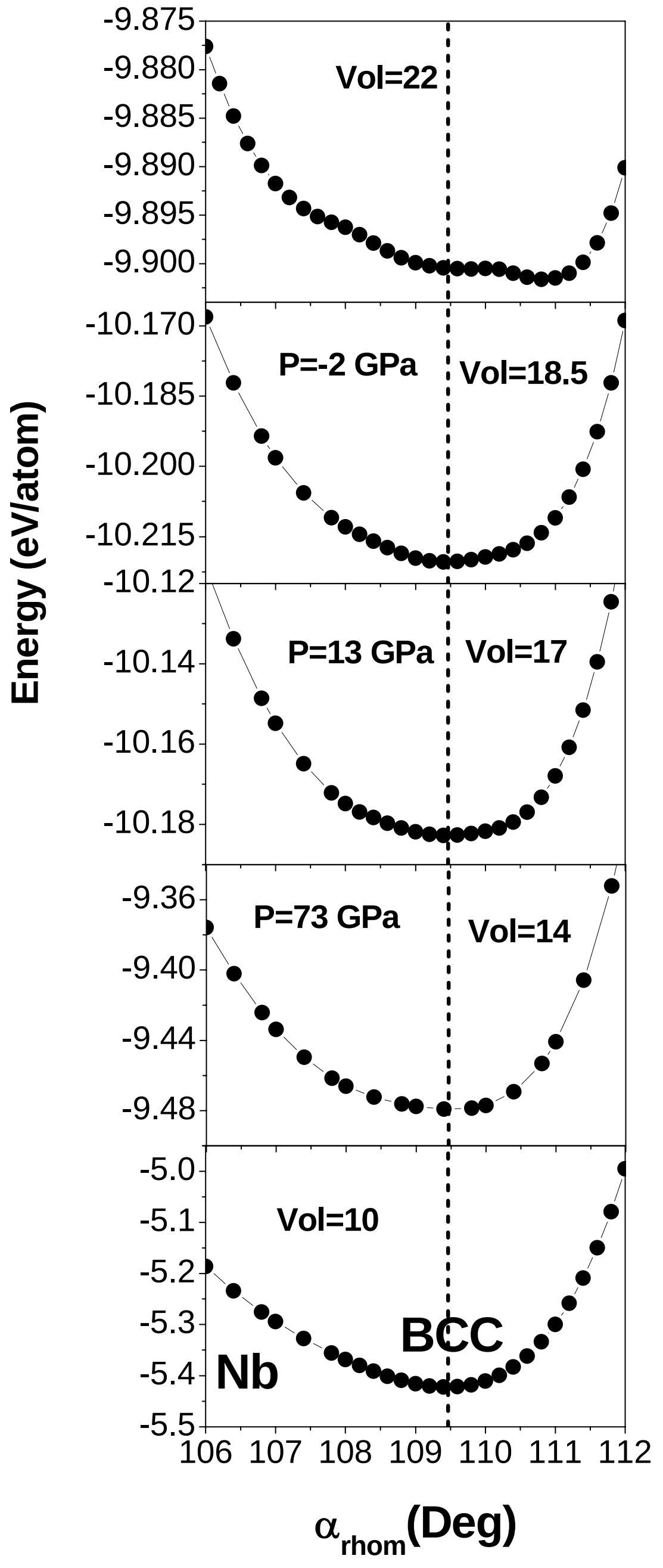

**Fig. 4a**

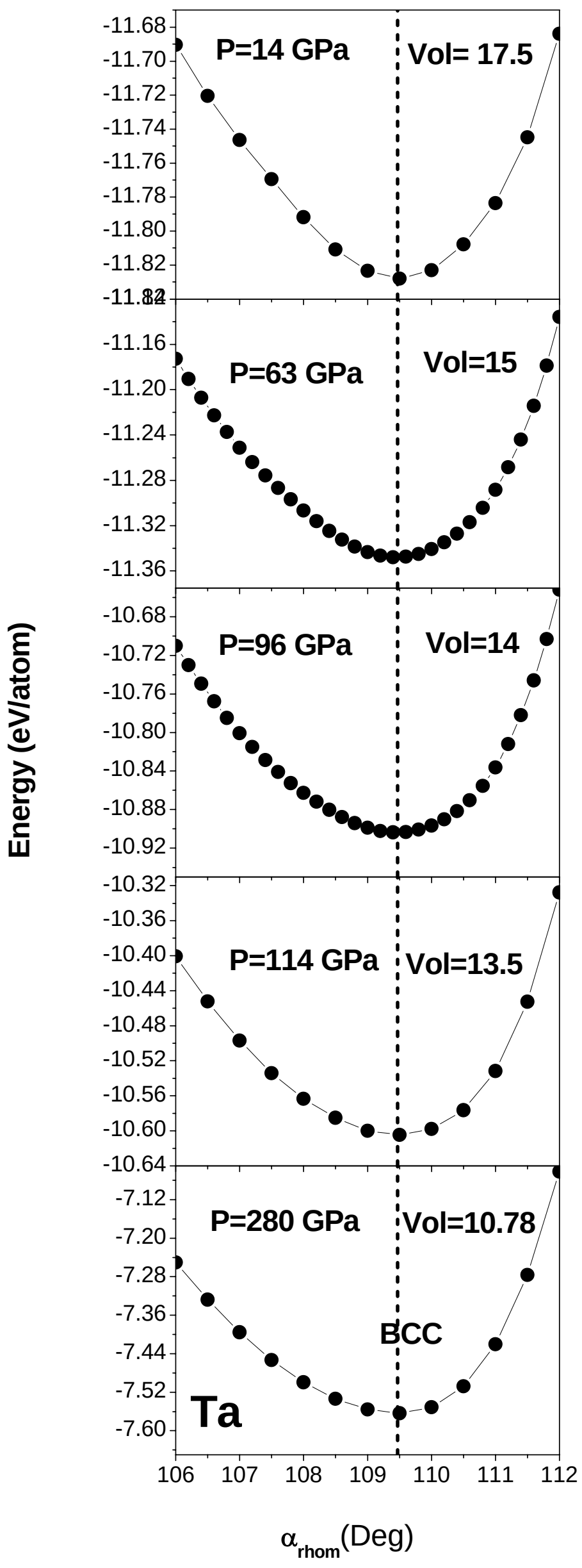

**Fig. 4b**

**Fig. 4 (a,b):** The calculated total energy variation with $\alpha_{rhom}$ at several volumes for Nb and Ta. The volumes given in figures are in units of $Å^3$ per atom.